\documentclass[pdflatex,sn-mathphys]{sn-jnl}

\jyear{2023}%

\theoremstyle{thmstyleone}%
\usepackage{listings}
\lstdefinestyle{tablesql}{
    language=SQL,
    frame=none,
        backgroundcolor=\color{backcolour},   
        commentstyle=\color{codegreen},
        keywordstyle=\color{codepurple},
        numberstyle=\footnotesize\color{codegray},
        stringstyle=\color{codepurple},
        basicstyle=\footnotesize,
        breakatwhitespace=false,         
        breaklines=true,                 
        captionpos=b,                    
        keepspaces=true,                 
        numbers=left,                    
        numbersep=-10pt,                  
        showspaces=false,                
        showstringspaces=false,
        showtabs=false, 
    breakatwhitespace=true,
    framexleftmargin=10pt,
    framexrightmargin=5pt,
    framexbottommargin=4pt,
    showstringspaces=false,    
}

\begin{document}

\title[Article Title]{Analytical Queries: A Comprehensive Survey}


\author*[1]{\fnm{Petr} \sur{Kurapov}}\email{petr.a.kurapov@intel.com}
\author*[2]{\fnm{Areg} \sur{Melik-Adamyan}\email{{areg.melik-adamyan@intel.com}}}



\abstract{Modern hardware heterogeneity brings efficiency and performance opportunities for analytical query processing. In presence of continuous data volume and complexity growth, bridging the gap between recent hardware advancements and the data processing tools ecosystem is paramount for improving the speed of ETL and model development. In this paper, we present a comprehensive overview of existing analytical query processing approaches as well as the use and design of systems that use heterogeneous hardware for the task. We then analyze state-of-the-art solutions and identify missing pieces. The last two chapters discuss the identified problems and present our view on how the ecosystem should evolve.}




\maketitle

\section{Introduction}

Historically, data analysis was performed using DBMSes.
The very first mainframes were developed for ``Data Banks'' processing.
Rapid technology development led to data being generated everywhere:
from industrial sensors telemetry to tracking personalized information on websites.
According to International Data Corporation (IDC~\cite{web:idc}), humanity produced 64 
zettabytes (1 ZB ~ 1.000.000 PB) of information in 2021,
and the number is expected to grow at an annual rate of 23\%.
Such a rate outpaces the performance improvement of computational power.
In addition to the volume growth, the structure of the data is getting more complex as well. 
Traditionally, this field is called “big data” and is characterized by three main points:
\begin{itemize}
    \item High volume,
    \item High speed of change,
    \item High complexity and diversity.
\end{itemize}

Big data aggregation and analysis generated demand for rising technologies
such as Hadoop~\cite{polato-hadoop}, Spark~\cite{spark}, and the Data Lakes concept~\cite{data-lake}.
Since the mid-2010s, the need for cheap storage and processing of
a large joint volume of structured, semi-structured, and unstructured data,
with the performance requirement of traditional DBMSes, has become relevant.
Such systems suffer from performance degradation over time as
the data volume grows and new acceleration methods are required.
Hardware acceleration, such as running a query on a GPU or
using different types of memory for intermediate storage,
provides a way of coping with the ever-growing performance
demand for data preprocessing. Integrating a hardware accelerator into a
DBMS for query processing requires reworking key mechanisms.
Because of the discrepancies in instruction set architectures between
CPUs and hardware accelerators, different types of queries may benefit from
different types of execution devices.
A direct comparison shows there is no ``best'' option for all the queries.

To address the lack of computation power, accelerator-based DBMSes
and particularly GPU-based DBMSes emerged.
However, in the presence of on-GPU memory restrictions and the nature of queries,
it became necessary to control the flow of execution and find the best distribution
of work between devices in the system. Query optimization is known to be
an NP-hard problem and the use of heterogeneous hardware increases the complexity and search space,
even more, thus requiring new approaches for query optimization and execution.
In addition, the complexity of administering is added.

\section{Analytical queries}\label{sec:analytical-queries}

As known, there are three main types of queries:
\begin{itemize}
    \item OLTP (online transaction processing) --- processing of single repetitive transactions,
    e.g., changing the state of an account in a bank through an ATM.
    This type of query is characterized by access to a small part of the data
    (updating one record), and constantly making changes, resulting in
    synchronization conflicts during parallel processing.
    \item OLAP (online analytical processing) --- analytical queries to identify additional
    information from existing data, such as quarterly revenue averages from sales of certain types of goods.
    This species is characterized by computationally intensive (a large number of data sources and tables,
    high nesting of joins, a large number and complexity of predicates) queries with access to
    a large part of the data, read-only access to data, and, as a result, query independence.
    \item HTAP (hybrid transaction/analytical processing)~\cite{htap} --- hybrid processing of small transactional
    changes (OLTP) and analytical queries (OLAP). Both patterns are characteristic of this species.
\end{itemize}

We focus on online analytical processing (OLAP). A typical OLAP query consists of multiple high-level steps:
\begin{itemize}
    \item Data search --- a nontrivial task for distributed systems (HDFS~\cite{hdfs}, S3~\cite{s3})
    or in the presence of multiple data sources.
    \item Data loading --- accessing disk hierarchies and network transfer.
    Data could be split and sharded between nodes in a cloud/cluster.
    \item Data validation and preprocessing --- running checks and transformations to match
    a format of an analysis tool.
    \item Data analysis and reporting.
\end{itemize}

Analytical queries are characterized by complex, long-running processes that access
high volumes of data from different sources. A query may generate multiple
sub-queries that access various services. For instance, in order to validate
a bank transaction a query may consider such parameters as the user’s location,
the number of requests made by that user, the mean number of requests
classified as fraud from a particular network point, etc.
Since the number of data sources is high in any real-case scenario,
aggregating data in a single data ``warehouse'' is necessary.
The process of gathering and preprocessing the data to store it in a data warehouse
is usually called ``Extract, Transform and Load'' (ETL) and is a part of OLAP.
OLAP, thus, covers most of the data processing scenarios in
data scientists and business analysts' day-to-day jobs.
Traditionally, data analytics was performed by the means of a DBMS.

\subsection{DBMSes for analytics}\label{subsec:dbms-for-analytics}
Relational DBMSes provide a rich set of capabilities for data access and processing optimization.
Abstracting away the layout and access algorithms implementation
a DBMS engine can control how the data is stored and optimize query execution plans.
Different types of queries to a single data warehouse can
cause performance degradation and lower the responsiveness of the system.
Transactional queries require synchronization mechanisms,
while much slower, analytical queries need data consistency throughout their execution.
For the two workload types to coexist specific techniques like
snapshots are necessary~\cite{htap}.
Splitting the two types of workloads is especially important for
recurring queries, for example, during a dashboard update.
A rigidly defined schema and a specialized storage layout
(e.g., physical duplication, columnar representation) help boost the performance
of computationally intensive analytical queries (the nature of an analytical query
implies accessing most of the accessible data, for example, the mean calculation requires accessing the whole column).
High data volume demands distributed storage and computing.
The complexity of queries additionally demands thorough query planning
in terms of the level of parallelism and algorithms chosen.

\subsection{Exploratory data analysis}\label{subsec:exploratory-data-analysis}
A typical data analysis scenario in searching for new patterns and insights differs from the recurring nature of dashboard updating.
Such a workload implies ad-hoc queries~\cite{seattle} and exploratory data analysis~\cite{tukey77}.
The random nature of such queries limits the opportunities for optimization.
In the dashboard creation case, query results may be cached and reused.
In the case of ad-hoc queries, the reuse of complete query results becomes impossible.
However, techniques for reusing more granular results
are still available~\cite{inmem-expr,plan-stitch,shared-wl-opt}.

\subsection{Example}\label{subsec:query-example}
\begingroup
\begin{lstlisting}[lastline=78,language={SQL},caption={TPC-H Q6},label={lst:tpch-q6}]
SELECT sum(l_extendedprice * l_discount) as revenue
FROM lineitem
WHERE l_shipdate >= date '2023-01-01'
    AND l_shipdate < date '2023-01-01' + interval '1' year
    AND l_discount between 0.06 - 0.01 AND 0.06 + 0.01
    AND l_quantity < 24;
\end{lstlisting}
\endgroup

Consider a typical analytical query from the TPC-H decision benchmark set (listing~\ref{lst:tpch-q6}).
The query forecasts revenue change given a discount policy for some company products.
The estimation is based on the ``lineitem'' table that contains pricing and shipping data.
There are three high-level query execution steps for running the query on a single node:
\begin{enumerate}
    \item Find and get access to lineitem’s data. For in-memory storage, that would mean acquiring some kind of a memory descriptor.
    \item Allocate memory for the result table. The size of the table can be estimated using precalculated
    statistics and reallocated during execution if needed. In this particular case, the table degenerates into scalar output.
    \item Loop through the table and calculate predicates for each tuple.
    If all predicates are fulfilled, put the result value into an empty result table slot (or apply an aggregation function).
    A more complex query would sequentially apply a set of operators to each tuple.
    Some of those operators may require a full intermediate results materialization in memory.
\end{enumerate}

Thus, query execution consists of data search, reading, predicates, and aggregations calculation, joining, sorting, and other operator evaluation.
Consequently, query execution optimization focuses on reducing the execution time of operators’ sequences.
Optimization can be aimed at reducing the execution time of a single query, increasing the throughput of the DBMS,
reducing the average response time to a query or the cost of executing a query in dollars, and other metrics.

Most modern DBMSes support columnar data organization and some vectorized processing or
query compilation techniques~\cite{neumann-2011} that make use of the data format to process several tuples at a time.

\subsection{In-memory DBMSes}\label{subsec:in-memory-dbms}
Query execution optimization is an active research topic for more than half a century.
Many optimization approaches were developed during that time ranging from architectural solutions to efficient algorithms development~\cite{volcano, arch-query-comp, distribute-join}.
The main goal of all those methods has been to increase hardware usage efficiency.
Optimization focus, however, changed over time due to rapid data volume increase and overall industry development~\cite{one-size-fits-all, main-mem-impl}.
Classical DBMSes pay the most attention to comparatively slow disk interactions.
Minimizing the access time has spawned many optimizations, for example, by data storage (compression),
by ways of accessing the data (indexing), by result accuracy (Bloom filters) etc. The ``producer/consumer'' (Volcano) model has emerged,
allowing keeping tuples in caches and registers as long as possible~\cite{volcano}.

Increased main memory capacities and lower prices drove the creation of in-memory databases~\cite{main-mem-impl}.
Storing data in DRAM changes the performance characteristics of algorithms developed for classical DBMSes.
Performance bottleneck shifts from memory access to data processing~\cite{where-time}.
It becomes profitable for the DBMS to be aware of the memory hierarchy and use mechanisms such as prefetching.
The use of secondary indexes is becoming less relevant~\cite{should-scan-or-probe}.
The number of instructions required to execute a query becomes an important factor,
along with the performance of the memory subsystem~\cite{neumann-2011}. Data access speed is now enough to
enable physical parallelism at the level of multiple cores and vector instructions~\cite{vector-vs-compile}.
Finally, the depletion of these techniques’ potential~\cite{vbench} has led the industry to
explore ways to increase efficiency through the use of dedicated hardware accelerators~\cite{adaptive-hetero}.

\subsection{Heterogeneous systems}\label{subsec:het-systems}
\begin{figure}
    \centering
    \includegraphics{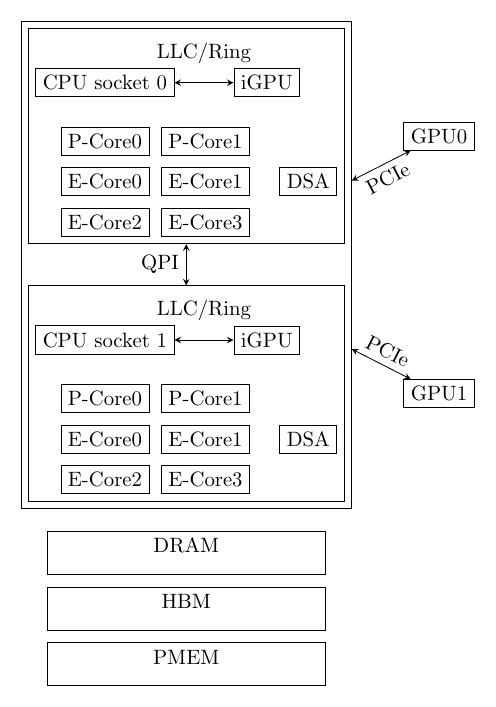}
    \caption{Heterogeneous node example}
    \label{fig:het-platform}
\end{figure}

By heterogeneity, we mean a possibility of joint query execution on two or more devices with different architectures.
Those could be two CPU cores that provide different capabilities such as vector extensions (fig.~\ref{fig:het-platform}) or
two completely different devices such as CPU and GPU, TPU, and other accelerators (examples may include Habana Gaudi, Ali-NPU, etc.).

Instruction set architecture (ISA) differences and microarchitectural features allow the hardware to
adapt to certain types of workloads by either conserving more power or minimizing the number of cycles
required to complete the task~\cite{hw-sw-stack-for-het}.
For example, a GPU would have a higher on-device memory throughput in
comparison to a CPU and would allow more concurrency exploitation possibilities.
However, due to the relative simplicity of each execution unit,
complex operations such as division and complex branching are less efficient.

Using multiple non-uniform devices in a single system requires supporting the target
code compilation and an appropriate runtime.
Programming models may differ from device to device.
There may be dependencies, like in the GPU case, in which execution management has to be handled on the CPU.

In addition to execution units, heterogeneity also occurs in the following characteristics of a system:
\begin{itemize}
    \item access to various types of RAM (DRAM, HBM, PMEM) with main parameters of access time, throughput, volume, and persistence;
    \item data location in the disk hierarchy (SSD, HDD) and NUMA;
    \item networking.
\end{itemize}

Applying heterogeneous systems to query execution can speed up particular relational operator
implementations~\cite{gpu-acc-analyt-engines, fpga-partitioning} as well as analytical
queries as a whole~\cite{het-aware-placement, hero-db}.

\subsection{Heterogeneous systems usage issues}\label{subsec:het-issues}
Using heterogeneous systems to further optimize queries is challenging~\cite{one-doesnt-fit-all}.
Changing the main execution device does not automatically give performance benefits for all query types.
On the contrary, a GPU loses to a CPU on a wide class of queries even if the data is local to the GPU~\cite{one-doesnt-fit-all}.
The dependency on workload type creates opportunities for optimization: which device to choose for a query or parts of the query.
High-speed networking in data centers with the new memory types represents a new class of latencies characterized by microseconds~\cite{killer-us}.
As the access time goes down the comparative computational throughput will decline.
Today’s methods of hiding memory latency rely on instruction-level parallelism and I/O ---
on operating system mechanisms such as context switching.
Both approaches do not scale well for operations in the microsecond range:
the instruction level parallelism is limited and insufficient,
and the context change is comparable in time~\cite{killer-us} to the transfer of data from non-local memory and will introduce significant delays.

Thus, to use a heterogeneous system to its full potential a system developer has to resolve the following issues:
\begin{enumerate}
    \item An efficient parallelism and vectorization model is required to utilize hardware capabilities.
    Devices’ programming models may differ.
    The query processing system architecture should take these differences into account and plan the query accordingly.
    \item Data transfer between devices introduces additional overhead.
    A mechanism for best memory and device mapping determining is needed for minimizing data movement expenses.
    \item Given all the heterogeneities the search space for the optimal query execution plan grows.
    The optimizer requires efficient search algorithms and new heuristics.
    \item Direct CPU cost models reusing is impossible due to architectural discrepancies.
    A new cost model is required that considers data remoteness and device features.
    The performance of a specific algorithm implementation is highly dependent on certain architectural and
    micro-architectural hardware properties.
    An efficient hardware utilization requires accurate query execution planning.
    \item With the increase in the number of platforms used,
    the support and development of an engine become more complicated.
    Complexity and code base size reduction methods are needed.
\end{enumerate}

Before diving into these issues, let's consider the fundamental classical mechanisms and approaches to query optimization.

\section{Query optimization}\label{sec:query-optimization}

\begin{figure}
    \centering
    \includegraphics[width=\textwidth]{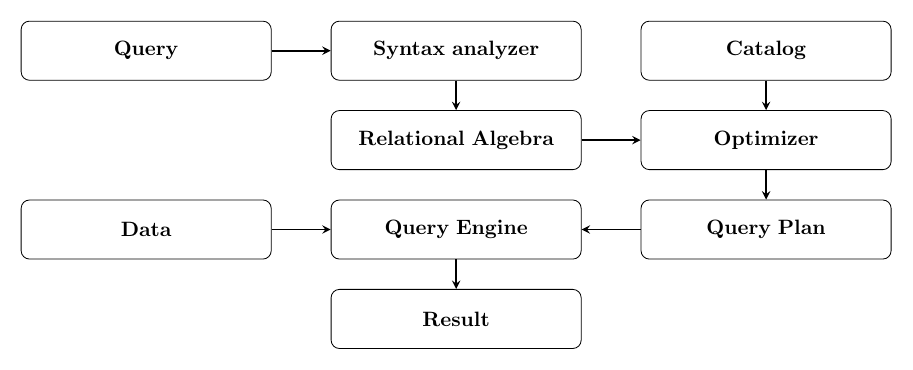}
    \caption{Schematic diagram of query execution.}
    \label{fig:query-exec}
\end{figure}

Target optimization parameters differ by workload type~\cite{one-size-fits-all}. It can be single query latency,
the overall throughput of queries per second~\cite{shared-wl-opt}, the number of processing nodes (servers, threads, cores),
memory footprint, power consumption, etc. For OLAP scenarios there are two main parameters:
\begin{itemize}
    \item Single query latency as an optimizer target. This parameter is important for any type of query.
    \item Execution cost as a resource quantity measure.
\end{itemize}

Analytical query optimization begins with logical query plan construction represented by a direct acyclic graph (DAG) of relational operators (Fig.~\ref{fig:query-exec}).
Query plan optimization consists of:
\begin{itemize}
    \item Finding efficient implementation (physical operator) for logical plan operators.
    Physical operators can either be implemented in the engine or generated during query execution.
    \item Choosing a data access strategy. This involves picking the right
    secondary indexes and choosing how to store intermediate results.
    \item Finding the optimal join ordering and join algorithms.
    \item Choosing a parallelization strategy.
\end{itemize}

The constructed physical plan is then passed to the execution layer. Physical plan execution performance is determined by the execution model.

\subsection{Execution model}\label{subsec:execution-model}
There are several popular query execution models:
\begin{itemize}
    \item Iterative model (volcano/pipeline/tuple-at-a-time)~\cite{volcano, volcano2} --- the classical model in which each relational operator
    implements a method compared to a CPU and each record.
    The execution engine traverses the graph of operators,
    calling the ``next'' method recursively until a leaf node returns some record.
    The record is then processed by a chain of operators until intermediate materialization is required (late materialization~\cite{posdb}).
    This way, records are kept in CPU registers and access to relatively slow disks is minimized.
    \item Operator-at-a-time or full materialization~\cite{pipe-coproc} --- each operator immediately materializes its results into the memory.
    Such an approach may work well for OLTP queries when intermediate tables are relatively small and can fit into caches.
    \item Vectorized processing (vector-at-a-time)~\cite{vectorwise} --- same as the iterative model,
    but instead of a single record the “next” method returns a batch of tuples.
    The pack of records can then be efficiently processed by vector execution units (SIMD).
    Can perform well in OLAP cases.
    \item Compiled processing (usually, operator-at-a-time or morsel driven~\cite{morsel}) --- implemented,
    for example, in Hekaton~\cite{hekaton}, MemSQL~\cite{memsql}, Spark~\cite{spark}, and Impala~\cite{impala}.
    There are three main approaches to applying compilation techniques:
    standalone expression compilation, query compilation, and automatic interpreter specialization.
    Usually, the compilation is implemented via some way of operators templates expansion~\cite{holistic-query-eval} or
    via building a syntax tree with consequent lowering to one or several lower-level intermediate representations (IR).
    Code generation quality depends on IR expressiveness,
    i.e. the ability to represent algorithms’ details for efficient execution on the hardware~\cite{arch-query-comp}.
    Generating source code or IR for a query allows applying all
    the classical compiler techniques such as function inlining or loop invariant code motion.
\end{itemize}

The volcano model as well as full materialization are inferior to the latter two
when it comes to OLAP performance due to significantly higher overhead for intermediate materializations~\cite{monetdb}.

A comprehensive comparison of compiled and vectorized processing models can be found in~\cite{vector-vs-compile}.
The results show that there is no single best option between compilation and
vectorization in terms of both performance and usability.
The vectorized model helps with profiling and debugging, but requires manual tuning of primitives.
Compilation abstracts low-level details away and allows using all the compiler optimizations.
The number of instructions in the resulting compiled query can be significantly less than
of what a vectorized engine can generate.
However, analyzing a compiled loop of consecutive operators is way more complex than following a sequence of vectorized primitives.

In the case of a heterogeneous system, vectorized execution can be inappropriate.
For example, the cost of GPU global memory access may exceed the benefit of vectorization ---
efficient operator execution may not require an additional parallelism layer (see~\ref{sec:parallelism}).
On the other hand, compiling large operator sequences increases register pressure
(a measure of the number of free registers) and can potentially cause high resource contention.

Automatic interpreter specialization offers interesting possibilities for
interaction between generated code and the engine’s binary.
For example, using JVM for dynamic query compilation allows using cross-optimization between
the query code and DMBS code~\cite{jvm-comp}.
GPU programming model, however, constraints such opportunities:
executing code on the GPU in interpret mode is highly inefficient.

The more complex the pipeline of operations, the more it is subject to
the so-called control flow divergence when using vector instructions,
that is, an increase in the number of idle SIMD channels
(to simulate parallel execution of vector branch instructions,
GPUs mask channels as inactive for which the predicate fails).
One of the promising techniques for compacting a divergent control flow is to
split operator pipelines into stages with partial intermediate materialization~\cite{vector-vs-compile}.
Small intermediate buffers allow previous operators’ results ``compression'' by
rearranging the values of the vector (SIMD) registers that turned out to be inactive,
and make it possible to effectively use the prefetch instructions.

Similar effects are observed when generating code for operator pipelines for GPUs ---
up to half of the execution time threads remain inactive~\cite{jit-gpu-simd-channels}.
Injecting additional pipeline splitting operators improves the efficiency of hardware use and,
as a result, engine performance.

One of the most important areas of optimization is the efficient use of the available multilevel
(multiprocessor, multithread, SIMD) parallelism provided by modern hardware.

\subsection{Parallelism}\label{sec:parallelism}
CPU and accelerators are developing along the path of increasing resources for parallel execution,
both in terms of instruction level parallelism (pipeline depth increase, out-of-order execution)
and vectorized computations.
Exploiting parallelism for analytical query processing is natural: each query is independent (“interquery”).
Moreover, the majority of operators can be parallelized (“intraquery”) to enhance resource utilization and reduce data access time.

The classical approach to induce parallelism into a query plan is to introduce
a new type of physical operator called the “Exchange” operator~\cite{volcano}.

\begin{figure}[ht]
    \centering
    \includegraphics{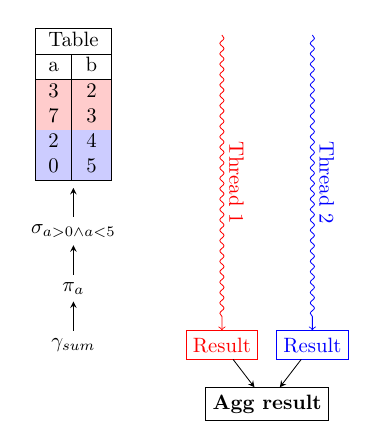}
    \caption{Parallelism types: Horizontal parallelism}
    \label{fig:hor-par}
\end{figure}

\begin{figure}[ht]
    \centering
    \includegraphics{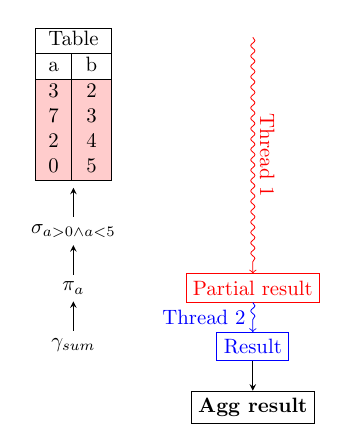}
    \caption{Parallelism types: Vertical parallelism}
    \label{fig:ver-par}
\end{figure}

There are two main types of intraquery parallelism:
\begin{itemize}
    \item Horizontal parallelism (Fig.~\ref{fig:hor-par}) --- each operator can run with multiple threads.
    For example, a scan-filter can split a table into partitions and run in
    parallel with each thread accessing its own partition.
    The obtained results then must be aggregated to preserve the uniformity
    of the operators’ interface. In this particular example, when executed on a CPU,
    an engine can set up a global counter in shared memory and update it
    atomically to find the position (i.e. calculate a prefix sum) in the final result buffer.
    A similar approach is used in the implementation of the join operator using hash tables~\cite{pipe-coproc,gpu-hash-join}.
    From the physical plan point of view, the control over data splitting and
    reduction is controlled by a class of exchange operators that do not change plan semantics.
    \item Vertical parallelism (Fig.~\ref{fig:ver-par}) --- in contrast to horizontal,
    uses the idea of pipelined execution of sequential statements (or independent ones) in a graph of the physical plan.
    If the next operator does not require the full materialization of the results of the previous one,
    then it can start execution without waiting for the child operator to finish ---
    similar to the producer/consumer model.
    Using a pipeline as a source of parallelism has scalability limitations:
    the pipeline’s length must be sufficient, and the operators must be balanced in terms of computational load.
    The idea finds its application in multi-core systems in a modification~\cite{morsel},
    which solves the scaling problem by dynamically balancing tasks with small
    data fragments (morsel-driven parallelism).
    Another use case for vertical parallelism is the parallel execution of data-independent operators.
\end{itemize}

In a distributed system, the time to transfer data from memory to an executor varies greatly,
so partitioning and locating data in a non-uniform memory access architecture (NUMA) becomes an important problem.
\cite{morsel,scan-numa}~show that explicit adaptive ``hot'' data placement control and
execution scheduling can significantly improve system throughput.
A ``morsel'' is a small fragment of data that can be processed by some query segment.
The result of the computation is materialized in memory.
The query planner splits a query into segments and assigns query pipelines’ computation to cores to avoid context-switching overhead.
To achieve a high level of parallelism, it is necessary that any query segment can work with morsels
(not only the input data and the materialized result of operators,
but also intermediate data structures, for example, a hash table).

The execution model is responsible for the performance of the query evaluation strategy.
With the same strategy, semantically equivalent plans will differ in performance.
A query execution engine needs to find the most efficient execution plan.

\subsection{Optimal plan search}\label{plansearch}
Modern cost-based optimizers originate from two projects:
IBM’s System R optimizer~\cite{systemr} improved in Starburst~\cite{starburst} (bottom-up approach)
and Exodus optimizer~\cite{exodus} (top-down).
The first approach uses dynamic programming to traverse the search space.
First, the cost of operations in the leaves of the search tree is estimated,
that is, table accesses cost (whether to use a secondary index and if so, which of the available ones).
For each leaf, all the available implementations are considered and the one with the least cost is chosen.
The resulting values are used to calculate the cost of a parent node, and so on up to the root.
In the second case, using cascades~\cite{cascade}, the optimizer will store the scanned hashed alternatives in a compact representation,
and be guided by the Bellman principle (any part of the optimal plan is optimal) going down the tree,
calculating the cost when the leaves are reached.

\subsubsection{Classical approach}
Due to the complexity of the task, optimization occurs in stages.
The classical optimizer consists of two stages~\cite{query-opt}:
\begin{enumerate}
    \item A set of predefined heuristic modifications are applied to the plan by a ``rewriter'' component.
    Examples of such rules may include: bringing the query to a canonical form,
    replacing data views with their actual definitions, and reducing the nesting of loops.
    \item Finding the optimal plan in the space of semantically equivalent ones.
    A cost model is used to compare a pair of plans. The optimizer estimates the operator
    execution cost using the statistics provided by metadata (table schemas and data statistics such as skews),
    as well as the size of the output in the case of the bottom-up approach.
\end{enumerate}

Since the number of equivalent plans when permuting join operators’ order turns out to be exponential,
in practice the enumeration space is often limited to subsets of trees of a certain type (for example, left-deep).
Such trees are even further pruned by heuristics.

The complexity of operation cost prediction models is limited by the query optimization time.
If the optimization time significantly exceeds the execution time and does not provide tangible benefits,
then it is more profitable for an engine to disable optimizations altogether.
However, as a rule, the optimizer allows for speeding up complex queries by an order of magnitude~\cite{how-good-qopt}.

Since the optimizer is based on an estimate of the operators’ cost and the output data sizes (cardinality),
accurate estimates of the statistical parameters of the current data are required (which, moreover, may become outdated).
Classical techniques are based on data assumptions: (1) independence, and (2) uniformity.
And therefore they can be wrong. With large errors in output data size estimation,
the cost model has almost no effect on the operator cost calculation accuracy~\cite{how-good-qopt}.
Therefore, there are many methods for improving these estimates.
An exhaustive review of the techniques is presented in~\cite{synopses}.
Here we briefly summarize the main approaches:
\begin{itemize}
    \item Histograms --- divide a table into groups based on a column value and store an estimate of the number of records for each group.
    \item Random samples --- a small subset of the table is duplicated and stored separately.
    To calculate predicate selectivity, the predicate is applied to the sample,
    and the result is extrapolated to the entire table under the assumption of data and sample uniformity.
    \item Wavelets --- the idea of applying a wavelet transform to a data set and storing its description as a set of coefficients.
    \item Sketches (FM, MinCount, LinearCount, LogLog, HyperLogLog)~\cite{sketeches} --- unlike random sampling,
    sketches are built over the entire data set and can give an accurate answer to
    a query about data characteristics (an example of the simplest sketch is the average value of a column).
    Combined with random sampling, accuracy improves~\cite{join-card-est}.
    \item Learned models --- apply machine learning algorithms to estimate the output size.
    A classic example of the approach for DB2 is described in~\cite{leo}.
    After 20 years, the idea of using learned models gives much more flexible and accurate results~\cite{dl-card-est}.
\end{itemize}

Adding one more device to the system at least doubles the number of options for each operator.
For the plan as a whole, this means exponential growth of the enumeration space.
Execution engines built for heterogeneous systems, thus, require new optimization approaches.

\subsubsection{Alternative approaches}
Organizing a search with heuristic reduction of the search space is not the only way to find the optimal plan.
For example, the idea of applying machine learning has been developed in the application of models not
only for estimating the statistical parameters of data but also as a way to solve the problem of estimating operators’ costs~\cite{learned-cost}
or optimizing the entire query~\cite{neo}.
One of the main problems of the trained models is the low quality of the generated plans at the ``zero knowledge'' stage.
NEO solves this problem by reusing the experience of optimizer developers to quickly train the model.
Another problem is the need to retrain the model when the statistical parameters of the data change:
models tend to adapt to the current data set. BAO~\cite{bao} solves this problem in a non-intrusive way.
Like NEO, BAO uses an existing plan optimizer and manages the process with ``hints''.
A convolutional model on a query plan graph reveals patterns and learns to choose the most beneficial set of hints.
Thus, errors of the heuristic optimizer are smoothed out.
The BAO approach is sensitive to the dimension of the search space, and with its significant increase,
the efficiency of the optimizer is not enough~\cite{bao-scope}.

Such training for plan optimization is built on top of query execution feedback.
Of course, the use of machine learning methods is not the only way to reuse historical information.
For example, in Microsoft SQL Server there is a plan optimization technique based on the idea of combining efficient parts of plans~\cite{plan-stitch}.
The technique requires solving two problems: (1) finding efficient sub-plans and (2) finding an efficient combination of these sub-plans.
For computationally-intensive non-repeating analytical queries, this approach is poorly applicable. 

Distributed systems require an additional layer of execution control.
For example, MemSQL~\cite{memsql} introduces an additional scheduling layer to control distributed nodes.
The query plan for a single node is extended with ``annotations'' that indicate the remoteness of the data being accessed by the operator.
At the execution stage, the annotation allows the DBMS instance (MemSQL organizes computing nodes in a shared-nothing way) to dynamically decide on the need to load data.

Another approach is to use randomized algorithms and metaheuristics.
For example, in PostgreSQL, the genetic optimizer~\cite{postgress-genetic} considers possible solutions to an optimization problem as a population.
Plan options are encoded with numeric strings that indicate the order in which the tables are joined.
The worst-valued individuals are replaced in the new population. New individuals are created using sections of strings (genes) with a known low-cost estimate.
The process continues until the optimizer evaluates a predetermined number of plans.
Unlike the standard PostgreSQL optimizer, the genetic algorithm allows it to reduce the number of iterated plans and efficiently search for the optimum for complex queries.

Thus, so far we identified 5 problems with using heterogeneous systems. When using classical methods for optimizing analytical queries, the following remain unresolved:
\begin{enumerate}
    \item The classic approach provides a concurrency model but needs to be extended to support multiple devices.
    \item There are methods for the efficient compilation of a query, such as~\cite{neumann-2011,morsel},
    but their adaptation for execution on different devices is necessary.
    \item Improvement of optimizer efficiency is required.
    \item A unified cost model is required.
    \item Methods to reduce the complexity of platform support are needed.
\end{enumerate}

\section{State-of-the-Art survey}
Over the past decade, research has identified two main approaches to integrating hardware accelerators into query execution engines:
using an accelerator to solve narrow problems such as plan cost estimation,
or full-fledged execution of relational operators on all available devices. 

The most common way to organize the interaction with an accelerator is to use the PCIe bus for data transfer and device control. In this configuration, to use specialized hardware as a database query accelerator, two factors limit the potential: limited device memory volume compared to main memory and relatively low data throughput between CPU and device. Given the limited speed of PCIe, the difference in memory bandwidth for the CPU and GPU does not allow the GPU to be used as a coprocessor that efficiently executes operator-level primitives. Regardless of the choice of an accelerator, for a "cold" start (that is when the data is not placed in the device's memory in advance), the data transfer time dominates the execution. With an increase in the transfer rate and a memory model that allows direct access (Direct Memory Access, DMA), the hardware’s efficiency increases. In addition to executing a query or certain statements on a GPU, other methods of using hardware accelerators have been explored in order to improve the performance of a DBMS. For example, accelerating the evaluation of predicate selectivity to improve the performance of the optimizer or bit-partitioning of tables and multi-stage execution.

The most common way to organize the interaction with an accelerator is to use the PCIe bus for data transfer and device control.
In this configuration, to use specialized hardware as a database query accelerator, two factors limit the potential~\cite{pump-volume}:
limited device memory volume compared to main memory and relatively low data throughput between CPU and device.
Given the limited speed of PCIe, the difference in memory bandwidth for the CPU and GPU does
not allow the GPU to be used as a coprocessor that efficiently executes operator-level primitives~\cite{cpu-gpu-fundamentals}.
Regardless of the choice of an accelerator, for a "cold" start (that is when the data is not placed in the device's memory in advance),
the data transfer time dominates the execution.
With an increase in the transfer rate and a memory model that allows direct access (Direct Memory Access, DMA),
the hardware’s efficiency increases~\cite{olap-hetero-analysis}.
In addition to executing a query or certain statements on a GPU,
other methods of using hardware accelerators have been explored in order to improve the performance of a DBMS.
For example, accelerating the evaluation of predicate selectivity to improve the performance of the optimizer~\cite{gpu-assisted-qopt}
or bitwise partitioning of tables and multi-stage execution~\cite{bit-partition}.

Device architecture differences complicate the optimization task. Basically, there are two ways to deal with the complexity:
separation of implementation and/or code generation into separate paths for each device,
or unification of code generation and optimizations.
The first approach makes it much easier to apply specific optimizations
(for example, the use of SIMD registers, managed shared memory or caches, multi-core, etc.) that reflect microarchitectural features.
The price of this convenience is the cost of maintaining the system.
Specialization for specific products creates barriers to system extensibility.
On the other hand, a single code generation path requires sufficiently expressive abstractions to be able to generate efficient code.

Next, we will look at the most significant existing approaches to using various heterogeneous platforms for query optimization.

\subsection{GDB}\label{gdb}
One of the first studies on the use of hardware accelerators showed~\cite{gpu-coproc-rel-op} that at least some of the relational operators
and their corresponding algorithms are advantageous to execute on or with a GPU in terms of performance.
However, to correctly determine how to divide the workload among the devices most efficiently existing cost models are not enough.

The purpose of the implemented GDB system was to compare the performance of analytical queries on the CPU and the GPU.
Therefore, each execution path was optimized separately for the device and in different programming languages,
taking into account architectural features.
To ensure the interchangeability of operator implementations and the possibility to execute the same operator together on different devices,
the execution in GDB was built on primitives. Because CPU and GPU primitives implementations differ,
the optimizer (which follows the model of~\cite{systemr}) requires an extension of the cost model to account for architectural differences and coordination between devices.
The model does not take memory latencies into account, relying on a large amount of data and the hardware's ability to hide latencies behind context-switching.
The memory access cost of each primitive is estimated analytically, taking into account the known data bus bandwidth in advance.
The total execution cost is composed of the data transfer time and the algorithm execution time.

In this model, the data of the coprocessor-executed operator was copied over the relatively slow PCIe bus every time it was called.
Due to this, the model showed low values of performance gain of the whole query despite the significant acceleration of individual operators.
In addition, creating a separate execution kernel for each operator has low GPU utilization potential~\cite{tpch-gpu-profiling}.
Nevertheless, with the right data placement strategy, the use of the GPU can improve performance and energy efficiency of execution~\cite{cpu-gpu-fundamentals}.

Another equally important problem with the approach is the duplication of code to support the coprocessor.
From the point of view of real system development, such architecture will significantly limit the developer's possibilities.

Thus, based on the first prototype of a heterogeneous DBMS three aspects of the problems outlined in paragraph~\ref{subsec:het-issues} can be distinguished:
\begin{enumerate}
    \item Duplication of algorithm implementation work for each operator (or lack of a single compilation path, aspect of problem~5).
    \item Low utilization of GPU resources in operator-at-a-time execution (problem~1).
    \item Low efficiency of interaction between CPU and GPU (problem~2).
\end{enumerate}

\subsection{Ocelot, HeroDB}\label{ocelot}
Ocelot~\cite{hardware-oblivious-paral} solves the problem of code duplication by using hardware-oblivious operators that are implemented in OpenCL~\cite{opencl}.
Compared with the base implementation of MonetDB~\cite{monetdb}, the authors show that the approach generates code on
different platforms that are comparable in performance to that default optimized implementation for
sufficiently large amounts of data (hides compilation and runtime overheads) and using a query plan that is initially optimized for execution on the CPU.

Approach advantages:
\begin{itemize}
    \item Portability. The implementation of operators in the OpenCL language allows the use of arbitrary hardware with support for the OpenCL interface.
    \item Ease of support. There is only one way to execute a query in the code base, so developers do not have to support many optimizations for specific architectures.
    Operator code generation relies on compiler optimizations provided by a vendor.
\end{itemize}

From a performance point of view, the approach has two main problems:
\begin{enumerate}
    \item The implementation of operators assumes the possibility of full data placement in the device memory.
    This rather artificial constraint prevents the model from taking advantage of hardware acceleration in the case of large input data sizes.
    \item Despite the laziness of computation, Ocelot has an "operator-at-a-time" execution model (see paragraph~\ref{subsec:execution-model}).
    This means that each operator, including those that are not a pipeline boundary (pipeline breaker), fully materializes the result in memory.
    Despite the high level of operator parallelism, this is not an optimal execution strategy (unnecessary materialization of intermediate results between non-pipeline-breaker operators).
    In addition, the advantage of the compilation method of execution is lost. User code (for example, predicates) becomes a separate operator instead of becoming part of an existing one.
    In other words, there are no full-fledged cross-operator optimizations.
\end{enumerate}

The model allows for making decisions about placing operators on the right devices dynamically.
Since execution on the accelerator is inevitably associated with overhead, depending on the primitive and the size of the input data in each case, the optimal placement of computations varies.
On simple queries, \cite{het-aware-placement}~shows how to use a simple learned model to determine the placement of operators to get a significant performance boost (up to 4 times),
compared with static placement on one of the types of devices (for example, running everything on a discrete GPU).
Experiments with parallelization of independent operators at the device level did not show any significant performance gain due to the high overhead in the experiment,
however, the model implies a potential gain from such a use case.

The performance gain from efficient use of all available resources is demonstrated by HeroDB~\cite{hero-db}.
By sharing the CPU and integrated graphics coprocessor, the prototype demonstrates an approximately twofold
increase in performance on random TPC-H queries with small data sizes compared to execution in single device mode (CPU or coprocessor).
It should be noted that the authors did not indicate the specific model of the processor used,
while the measurements~\cite{het-aware-placement} clearly demonstrate the expected effect of reducing the operating frequencies of
the processor and graphics cores in the joint execution mode due to the sharing of resources (an integrated graphics processor was used for the experiments).
In addition, the prototype uses two different implementations of operators that are specific to the execution devices,
which greatly complicates the support and development of such a system, although it allows using all available optimizations.

In~\cite{het-aware-placement}, to select the execution device, the optimizer uses a linear extrapolation (by the size of the input data) of the
accumulated historical performance data to estimate the execution time of the operator and a heuristic that maximizes device utilization.
This approach works in the case of statement-by-statement execution and is poorly applicable in the case of a joint
compilation of several operators since the space of execution options increases (which means a significant decrease in the amount of experimental data for extrapolation).
In addition to this shortcoming, the model is limited in choice if the operator's input does not fit in the device's memory.
Thus, the optimizer loses the potential gain from using the optimal device.

\subsection{Hawk, CoGaDB, Hype, HorseQC}\label{horseqc}
These systems use source code generation of the query in a language that can be executed on various platforms.
Such an approach still allows for having a single implementation for all devices, but at the same time potentially allows for generating code for several operators at once.
HAWK generates device-independent OpenCL code for pipelines (pipeline programs) and uses existing implementations of machine code generators for the desired platform (OpenCL allows HAWK to execute queries both on the CPU and on the GPU and other multi-core architectures such as Xeon Phi).
In contrast to operator-by-operator execution, this approach saves resources for intermediate results materialization between operators that are not the boundary of the pipeline, passing the values of records in registers.
Moreover, merging multiple statements into a single translation unit allows the compiler to apply standard optimizations.

An earlier optimizer [77] implemented two strategies: greedy (using the optimal logical plan and then inserting data copy statements to transfer data to another device) and brute force with a limit on the number of transitions between devices.
However, as noted later, simple heuristic methods are not enough even to solve the problem of choosing the optimal execution device [78]. HAWK, in order to improve performance, specializes code generation in terms of parallelization strategy, and memory access methods, and chooses the most profitable implementation of algorithms (for example, hash tables).
Device-specific modifications are then applied to the intermediate representation of the pipeline. The result of the optimizer's work is a substitution of the optimal modification.
Despite this, as the measurements show and the authors themselves note, the overhead of source code generation (using the entire compiler stack) significantly slows down execution.
The optimization in this approach is local in relation to the query pipelines and does not take into account the possibility of accelerating inter-operator interaction.
Also, the framework is not designed to take advantage of device-level parallelism.

HorseQC allows for gluing operator kernels together to minimize launching and memory copying overhead for each operator.

Combining source code generation and gluing operator kernels solves the problems of the single way to generate query code for different devices and its efficiency (problems 2 and 5).
However, due to high overhead, the approach cannot be used in real scenarios.
Reducing the level of abstraction and generating a near-native representation makes the approach more attractive.
A significant reduction in the complexity of the intermediate representation can reduce compilation time and increase query response time on relatively small input data sizes.
Flounder IR is an example of such a technique.
As the size of the data being processed increases, however, the advantage in compilation speed is leveled, since the execution itself becomes the main component of the query execution time.
In addition, closeness to machine code implies a rigid binding to the architecture of the execution device.
Another approach to reducing compilation time is the classical JIT approach with various stages of optimizations [80; 81]: interpretation, compilation without optimizations, and compilation with optimizations.
Increasing the optimization level for computationally intensive queries solves the compile time problem and has no problem with increasing data size, however, the technique is not suitable for every device.
The data transfer time in interpretation mode over a relatively slow bus loses compilation much more.

\subsection{Voodoo}\label{voodoo}
A less expensive way to solve the same problem of high overhead for generating an intermediate representation and compiling is to generate an appropriate intermediate representation.
The Voodoo~\cite{voodoo} algebra allows translating the graph of relational operators into an abstract intermediate algebra,
from which the binary code of the required device is obtained using Just-In-Time (JIT) compilation (the OpenCL driver and compiler were used for GPU code generation).
The algebra is a set of vector operators without state preservation (stateless).
As in~\cite{neumann-2011}, the query plan traversal forms "fragments" with intermediate representation forming directed graphs.
Each fragment is characterized by two parameters: the degree of data parallelism (which can be expressed in terms of OpenCL iteration space dimension) and the number of consecutive operators.
Operators with the same degree of parallelism (Extent) can be executed within a single tight loop, i.e. without expensive materialization of the intermediate result.
Proprietary algebra requires a separate mechanism to lower the abstraction level into a corresponding representation.
Unlike, for example, the intermediate representation of a compiler (LLVM),
a separate intermediate representation does not allow to use of an existing JIT compiler directly.
It requires a separate machine-dependent optimization module.
The process of typing a fragment by data parallelism degrees is redundant in most cases.
The optimizer knows in advance which operators should be combined into one fragment, and the implementations of the algorithms are known in advance.
This information should be enough to split the execution into kernels in terms of OpenCL.

\subsection{LegoBase}\label{legobase}
Another way of high-level representation and use of DBMS statement primitives is to embed them into a language and use JIT compilation~\cite{legobase}(LegoBase).
The idea is to enable DBMS developers to use high-level language (Scala in this case) to describe optimizations in a generative programming way.
Most of the Scala language constructs are well expressed using the concepts of the much more efficient C language.
Therefore, LegoBase extends the set of operations of the internal representation of the JIT compiler (LSM~\cite{lsm}) and generates efficient C code, which is then compiled.
LegoBase does not use these operators with a classical optimizer but applies the technique immediately after obtaining the physical plan.
The authors introduce a concept of "abstraction without regret" and show how their approach enables a high-performance query engine without the time-consuming use of low-level primitives.
They identify the major problems with previous solutions:
\begin{itemize}
    \item Expanding templates does not allow efficient use of cross-operator optimizations.
    \item Templates are hard to implement because they are low-level and there is no strict typing (partially true even for generating LLVM representation).
    \item The query compiler cannot use optimizations to inline query code in the DBMS itself since its scope is limited to the query.
    \item LLVM and similar allow to apply runtime optimizations (e.g. JIT), but they do not allow to use of high-level information (predicate selectivity).
\end{itemize}

Most of the arguments against the use of LLVM IR point only to the fact that the first versions of the optimizer~\cite{neumann-2011} did not have effective support for computing predicates and dynamic implementations of data structures.
Instead, precompiled modules were used. It should be noted that despite the authors using a 10-year-old compiler,
no significant increase in the performance of the generated code should be expected when upgrading to newer versions (a 15-20\% gain over 10 years).
That is, the compilation time will mainly suffer from migration to newer versions (approximately by a factor of 2).
Nevertheless, the solution has two important advantages, the ideas of which are reused in new solutions.
\begin{enumerate}
    \item High-level representation of operations with complex data structures, such as a hash table, allows for convenient management of the execution model and implementation details.
    This reduces the system support's complexity and increases the implementations' variability to improve performance depending on dynamic conditions.
    In particular, it allows LegoBase to conveniently change data representation (column store and row store).
    \item Using a unified JIT compilation system for both the queries and the DBMS source code opens up additional cross-functional optimization possibilities.
\end{enumerate}

Another approach is to introduce a new parallel intermediate representation.
Through such a representation, Weld~\cite{weld} (a common runtime library for NumPy, Spark SQL, TensorFlow, etc.) provides a minimal set of primitives for library communication, namely, for data movement and concurrency expression.

\subsection{HetExchange}\label{hetexchange}
HetExchange~\cite{hetexchange} extends the set of physical plan operators by adding three classes of specialized ones:
context switching between execution devices, data movement, and the data flow control operator.
With standard operators unchanged, such a framework allows for generating plans that explicitly control all available parallelism and achieve performance gains both relative to other homogeneous parallel executors, and a device-specific implementation (CPU or GPU).

Query scheduling is orthogonal to the execution model (vectorization vs. code generation), and according to the authors, the approach can be applied in any of the two models.
An unresolved problem remains the definition of a method for separating data and load in a heterogeneous system, that is, a cost model is needed that allows setting the parameters for distributing data between devices.

\subsection{VOILA}\label{voila}
Another dimension of search space expansion is offered by VOILA~\cite{voila}.
VOILA offers a language that abstracts the implementation of a query engine with comparable performance to real systems, made according to a given design.
Unlike the intermediate representations of Voodoo~\cite{voodoo}, Weld~\cite{weld}, and the like,
VOILA offers an abstract operator description and an additional FUJI machine code generation level that allows for describing algorithmic details and variating them to create new query engine implementations.
To describe new kinds of hash-joins with voodoo, for example, it is necessary to provide new operator implementations.
VOILA breaks down such complex tasks into components: hash calculation, bucket lookup, key predicate validation, moving to the next bucket, and so on.
Since the task of parallelizing scalar code is too difficult, each primitive always works with a vector representation by default.
To implement the scalar execution strategy, the size of the vector is set to 1, and for the column-at-a-time approach, to infinity.
For each query pipeline, VOILA can apply the most appropriate execution method, as if there were a query engine with exactly those characteristics.
VOILA also allows you to vary the way it executes within each pipeline with a special Blend expression.
By blending expressions, the engine defines a new implementation (variant) inside an arbitrary scope (some fragment of the code that evaluates the query).
The data is buffered at the boundaries of such areas, and the subroutine is executed in a new variant.
This makes it possible to synthesize code according to models such as relaxed operator fusion (ROF)~\cite{rof}.
Such flexibility helps to find optimal implementations of algorithms from the execution model point of view and to get a significant increase in performance.

The price of such flexibility is the high cost of finding the right execution model.
In addition to enumerating the implementations of operators in the classical model, VOILA adds new dimensions within each operator.
Such an enumeration space required the authors to introduce restrictions even for TPC-H queries.

\subsection{OmniDB}\label{omnidb}
OmniDB~\cite{omnidb} attempts to solve the problem of choosing the optimal distribution of data across devices, given the requirements for efficient development and maintenance of a codebase for multiple architectures.
Since most of the system components specialized for different architectures can be reused, the authors propose a multi-level architecture based on~\cite{gpu-coproc-rel-op}.
To address portability and efficiency issues, extensible parallel kernels (query primitives, "qKernel") compute the query using adapters that hide all the details of the execution engine's architecture.
The kernel, in turn, is based on the Abstract Parallel Computing Model (OpenCL) and does not change depending on the device.
The collaborative (CPU/GPU) architecture model introduces N-threaded processors (Parallel Processing Element, PPE), each of which has its own memory space with block data access. 
This space may or may not overlap with the spaces of other processing elements, which allows considering the CPU or GPU as a whole PPE.
Kernel configuration for specific compute properties, such as cache sizes, is decided at the adapter level.
At the top level, the scheduler creates kernels for query operators and divides work, such as for processing a block of table tuples, between PPEs.
A queue (FIFO) is used to schedule blocks of work. The CPU and GPU are treated as PPEs based on their capabilities: the scheduler balances the load and tries to choose the most advantageous execution device in terms of performance.
The optimization uses a greedy algorithm:
\begin{itemize}
    \item The throughput of each PPE is estimated
    \item The current workload of the device is estimated, that is, how many blocks of work are waiting for execution (``pending'' status)
    \item Based on the collected data, the PPE with the highest throughput is selected, which is included in the set of underutilized (load level is determined by a threshold value) 
\end{itemize}

The cost model is based on two capabilities of adapters: to estimate the number of memory blocks accessed by the kernel and the number of instructions used. For I/O, the standard estimation method~\cite{dbms-book} is used.
The adapters then calibrate the size and latency of access to caches, the bandwidth of the PCIe bus, and the size of the data in the block (at different levels: query, operator, OpenCL kernel).

The greedy algorithm improves the efficiency of query scheduling in a heterogeneous system (up to 30\% more efficient than the naive distribution of work between devices), and architecture specialization reduces the number of second-level cache misses by 20\%.
However, the prototype has not yet shown a significant increase in performance. In addition, the locality of the decisions made does not allow the optimizer to achieve optimal efficiency.

\subsection{HeavyDB (OmniSciDB, MapD)}\label{mapd}
The in-memory DBMS HeavyDB is capable of executing queries on GPU and CPU and combines many of the research advantages described above.
The project uses Calcite~\cite{calcite} to generate logical query plans. The logical plan is optimized and represented using an internal expression algebra (abstract syntax tree, AST).
Expressions store query details, such as predicates. After generating the physical plan, HeavyDB translates the operators' pipelines ("step") of the plan into the intermediate representation of the compiler (LLVM) and generates code for each step for the selected platform.
The resulting kernel is executed on a columnar data representation (column-at-a-time). The uniformity of the principle of execution on the CPU (execution relies on JIT) and the GPU simplifies the architecture of the DBMS.
Using this mechanism, HeavyDB glues operators into a single kernel and applies cross-operator compiler optimizations (that is, it also adapts the technique~\cite{neumann-2011} for GPUs).
An important advantage of HeavyDB is that the DBMS minimizes the overhead of transferring data from the CPU to the graphics accelerator by storing hot data in the memory of the graphics accelerator between calls to step kernels.
This reduces the load on the data bus between the GPU and the CPU.

The main disadvantages of the approach are memory limitation (HeavyDB does not allow executing the query in the described scenario if there is not enough device memory for intermediate results) and the lack of joint execution of steps on different devices.
Some query types are not supported on GPUs, however, the DBMS provides a recovery mechanism (restarting part of the calculations on the CPU).
If an execution error occurs on the GPU the query will always be completed regardless.

\subsection{Centaur and Glacier}\label{centaur}
Heterogeneous execution is not limited to GPUs. Because of the great potential of field-programmable gate arrays (FPGA), there are many potential uses for optimizing query execution.

A framework for integrating with MonetDB and sharing an FPGA Centaur~\cite{centaur} places a pipeline of operators on cells and connects them with queues (FIFO).
Each supported operator is assigned to a hardware module, and the engine manages the placement of special threads and maps the thread to the user-defined function and the hardware module.
The result of an operator execution is written to the shared memory, and the CPU periodically polls the devices about their readiness.

Glacier~\cite{streams-on-wires, glacier} compiles the relational plan into a VHDL hardware description using the FPGA primitive library.
The description is passed to the synthesizer to generate the FPGA configuration. Each tuple of N bits is represented as a bus of the same width.
The primitive pipeline on the FPGA can process a stream of tuples one per clock.

\section{Open problems}
Existing approaches offer good mechanisms for exploiting parallelism (problem 1). In particular, the most promising approach is that described in section~\ref{hetexchange}. The approach scales well, but at the same time does not expand the search space as much as discussed in section~\ref{voila}. Joint compilation of pipelines like in HorseQC (Section~\ref{horseqc}) can significantly reduce the overhead of data movement (problem 2). The compilation itself, however, is not enough. Efficient execution still requires precise query planning (problem 3).

The methods proposed in sections~\ref{omnidb} and~\ref{horseqc} rely on simple heuristics and a greedy algorithm. This is not enough for an efficient optimizer. The cost model discussed in paragraph~\ref{gdb} estimates the cost of isolated operators. To use pipeline compilation, we need to extend it (Issue 4). Finally, a unified programming model for heterogeneous systems discussed in~\ref{ocelot}, reduces the complexity of development (problem 5). The disadvantage of this solution is the higher compilation time and lower expressiveness of the intermediate representation (partially solved with the transition to the generation of the intermediate representation).

Thus, the following problems remain unresolved:
\begin{itemize}
    \item Efficient search for the optimal plan in the extended search space.
    \item Extending the cost model to account for joint compilation of operators (pipelines).
    \item Increasing the speed of a single code generation path for heterogeneous platforms.
\end{itemize}

In addition to these aspects, we mentioned the problems of distributed scheduling in Section~\ref{plansearch}. The central task remains to quickly find effective plans and expand the cost model. This task is relevant for the most common platforms consisting of a CPU and a GPU due to their high prevalence. An integrated GPU is found in most modern desktop processors, and most servers provide the ability to connect powerful general-purpose discrete GPUs (GPGPU). And a significant part of the operators (for example, the calculation of complex predicates) can be effectively executed on the parallel architecture of GPUs.

\subsection{Discussion}
The current tendency of hardware development leans on specialization as a major driving force for performance improvement.
Data science tooling, thus, needs to adapt and provide ways of hiding the complexity from users who should be concerned with the actual problem rather than the performance optimization of data analytics pipelines.
That means a platform should automatically pick ``the right tool for the job''. Given the high dimensionality of the search space the task at hand proves challenging.
There are numerous ways for using available hardware accelerators, types of memory, and networking. On the other side, there are plentiful tools for different kinds of tasks such as specific machine learning algorithms (Data Ingress, Classical Machine Learning, Natural Language Processing, Computer Vision and image processing, Deep Learning, AutoML, etc.), feature engineering tools and storage, model management, experiments tracking, and more. In-house enterprise solutions are built on top of a set of tools from the ecosystem and have to integrate, maintain and monitor all the components. Plug-and-play components that provide out-of-the-box higher performance or efficiency are hence of great importance.

Improving the efficiency of a data processing engine as any other system fundamentally relies on two factors: ``data'' and ``compute''.
The ``data'' factor can be affected by system bandwidth and latency (networking and hardware capabilities, for example, PCIe throughput, or accelerator support such as Intel DSA~\cite{dsa}). The ``compute'' part is related to the hardware's raw computation power (includes ISA extensions such as vector instructions and different types of computation accelerators). Balancing the two factors in a system is crucial for it to be efficient. In heterogeneous environments, the ratio can vary. Controlling the required properties of data access (pattern, latency, persistence) and the type of computational engine unleashes opportunities for efficient computation.

Today's server nodes already may provide multi-socket CPUs with different types of memory and a set of discrete and integrated GPUs as well as additional accelerators. Exposing different sets of features to a workload creates a highly variable environment. The software stack must be robust and agile enough to accommodate itself to the change and use all the exposed capabilities to their potential. The volatile nature of components and interfaces asks for solid approaches to configuration and adaptation mechanisms.

And an additional factor entering the market is based on the new capabilities of interconnect, e.g. Compute Express Link (CXL~\cite{cxl}), which allows connecting physically decoupled ``compute'' into logically coherent super-``compute'' units. 

\section{Our vision}
After reviewing the existing work and pinpointing the most important gaps we arrive at a model implementation for a query engine. First, the focus of an analytical engine should be solely on analytical query processing.

In our opinion, a cost model-based approach partially resolves the planning issue as long as the model can accommodate new architectures. Such a solution can be implemented using high-level abstractions and cross-platform languages, the two most well-known of which are OpenCL C and SYCL. Single-source algorithm implementation minimizes development and support efforts.

The dynamic nature of the problem makes the data-centric code generation approach an appealing alternative. Optimization complexity can and should be hidden in vendor-provided compiler tools. A modular design follows the separation of concerns principle and narrows the optimization scope at each layer of the system. 
If the appropriate abstraction layers are in place, flexibility loss can and will eventually be minimized.

The code generation technique, as mentioned before, imposes a challenge to modeling a query performance behavior. The key observation for creating a robust model is that the set of operations in an analytical query processing engine has a limited known set of available operations. Those operations have distinct behavior on each platform that can be quantified by means of characteristic micro-benchmarking unified with comprehensive platform information including memory bandwidth, cache sizes, number of processing units, and so on.

We envision the emergence of modular, unified frameworks a la LLVM, where the logical plan, physical plan, and cost model will serve as common intermediate representations (IR) and different modules will work on these IRs to achieve the best problem-oriented solution. We already see small steps in that direction --- projects like Substrait~\cite{substrait}, Arrow, and Heterogeneous Data Kernels~\cite{hdk} bring the first bricks to the foundation of the future modular framework.

\bibliography{survey}


\end{document}